\documentclass[useAMS,usenatbib]{mn2e}
\usepackage{graphicx,subfigure,amssymb,times}
\usepackage[normalem]{ulem}

\newcommand{\RM}{\mathrm{RM}} 
\newcommand{\DFD}{\Delta F_{D}} 
\newcommand{\DF}{\Delta F} 
\newcommand{\cP}{\mathcal{P}} 
\newcommand{\cG}{\mathcal{G}} 
\newcommand{\er}{\epsilon_{\rho}} 
\newcommand{\lam}{\lambda} 
\newcommand{\Lc}{\mathcal{L}} 
\newcommand{\M}{\mathcal{M}} 
\newcommand{\ps}{\Psi} 
\newcommand{\pso}{\psi_{0}} 
\newcommand{\po}{p_0} 
\newcommand{\nel}{n_\mathrm{e}} 
\newcommand{\rad}{\mathrm{rad}} 
\newcommand{\sign}{\mathop{\mathrm{sign}}}

\newcommand\sfrac[2]{{\textstyle{\frac{#1}{#2}}}}

\newcommand\meanb[1]{\langle{#1}\rangle}
\newcommand{\dif}{\mathrm d}
\newcommand{\one}{_\mathrm{1}} 
\newcommand{\two}{_\mathrm{2}} 

\newcommand{\cm}{\,\mathrm{cm}}

\newcommand{\m}{\,\mathrm{m}}

\newcommand{\me}{\,\mathrm{e}}

\newcommand{\mi}{\,\mathrm{i}}

\newcommand{\kms}{\,\mathrm{km\,s^{-1}}}

\newcommand{\kpc}{\,\mathrm{kpc}}

\newcommand{\mkG}{\,\mu\mathrm{G}}

\newcommand{\p}{\,\mathrm{pc}}
\newcommand{\radmm}{\, \mathrm{rad}\,\mathrm{m}^{-2}} 

\newcommand{\yr}{\,\mathrm{yr}}

\title{Canals in Milky Way radio polarization maps}
\author[A. Fletcher and A. Shukurov]{A. Fletcher\thanks{E-mail:
andrew.fletcher@ncl.ac.uk; anvar.shukurov@ncl.ac.uk}  and A. Shukurov\\
School of Mathematics \& Statistics, University of Newcastle, Newcastle upon Tyne, NE1 7RU, U.K.}

\begin{document}

\date{Accepted . Received ; in original form}

\pagerange{\pageref{firstpage}--\pageref{lastpage}} \pubyear{2005}

\maketitle

\label{firstpage}

\begin{abstract}
Narrow depolarized canals are common in maps of the polarized synchrotron
emission of the Milky Way. Two physical effects that can produce these canals
have been identified: the presence of Faraday rotation measure ($\RM$) gradients
in a foreground screen and the cumulative cancellation of polarization known as
differential Faraday rotation. We show that the behaviour of the Stokes
parameters $Q$ and $U$ in the vicinity of a canal can be used to identify its
origin. In the case of canals produced by a Faraday screen we demonstrate that,
if the polarization angle changes by $90\degr$ across the canal, as is observed
in all fields to-date, the gradients in $\RM$ must be discontinuous. Shocks are
an obvious source of such discontinuities and we derive a relation of the
expected mean separation of canals to the abundance and Mach number of supernova
driven shocks, and compare this with recent observations by \citet{Haverkorn03}.
We also predict the existence of less common canals with polarization angle
changes other than $90\degr$. Differential Faraday rotation can produce canals
in a uniform magneto-ionic medium, but as the emitting layer becomes less
uniform the canals will disappear. We show that for moderate differences in
emissivity in a two-layer medium, of up to $1/2$, and for Faraday depth 
fluctuations of standard deviation $\lesssim 1\,\mathrm{rad}$, canals produced by
differential rotation will still be visible.
\end{abstract}


\section{Introduction}
Polarized (synchrotron) radio emission is a rich source of information about the
relativistic and thermal plasmas and magnetic fields in the interstellar medium
(ISM). Recent observations have revealed an abundance of unexpected features
that arise from the propagation of the emission through the turbulent ISM
\citep{Wieringa93, Uyaniker98, Duncan99, Gray99, Haverkorn00, Gaensler01,
Wolleben06}. Arguably the most eye-catching is the pattern of depolarized
canals: a random network of dark narrow regions, clearly visible against a
bright polarized background. These canals evidently carry information about the
ISM, but it is still not quite clear how this information can be extracted. Two
theories for the origin of the canals have been proposed; both attribute the
canals to the effects of Faraday rotation, but one invokes steep gradients of
the Faraday rotation measure ($\RM$) across the telescope beam, in a
Faraday screen \citep*{Haverkorn00, Haverkorn04}, whereas the other relies on
the line-of-sight effects producing differential Faraday rotation \citep{Beck99,
Shukurov03}. In order to use the canal properties to derive parameters of the
ISM, one must correctly identify their origin. 

In this letter we briefly discuss a few important aspects of the two theories
describing the origin of canals. Detailed discussion of relevant depolarization
mechanisms is presented in \citet[][and in preparation]{Fletcher06a}. Canals
produced by Faraday rotation measure gradients are discussed in
Section~\ref{sec:screen} where we show that these canals require a
\emph{discontinuous\/} distribution of free electrons and/or magnetic field; we
further suggest an interpretation of the discontinuities in terms of
interstellar shocks. The case of differential Faraday rotation is discussed in
Section~\ref{sec:diffrot}; in Section~\ref{sec:origin} we suggest an
observational test that can be used to identify the specific mechanism that
produces a given canal.

The defining features of the canals are as follows:
\begin{enumerate}
\item \label{itemi} the observed polarized emission $P$
approaches the polarization noise level $\sigma_P$, $P\la\sigma_P$;
\item the canal is about one beam wide;
\item the canal passes through a region of significant polarized intensity, say
$P\ga3\sigma_P$;
\item the canal is not related to any structure in total intensity, and so cannot be
readily explained by, e.g., an intervening gas or magnetic filament.
\end{enumerate}
Polarized intensity vanishes when emission within the telescope beam consists of
two equal parts with mutually orthogonal polarization planes. Thus, feature
\ref{itemi} most often arises because the polarization angle $\ps$ changes by
$90\degr$ across the canal. However, in Section~\ref{sec:screen} we predict a
new type of canal across which the observed polarization angle does not change.

We only consider canals occurring in properly calibrated maps:
\citet{Haverkorn04} argue that the observations we discuss in
Section~\ref{sec:meansep} do not suffer from missing large-scale structure;
\citet{Reich06} discusses the calibration of radio polarization maps in depth.

Polarized radiation is commonly described in terms of the complex
polarization,
\begin{equation}
\cP = p\exp{(2\mi\ps)}\;,
\label{eq:compp}
\end{equation}
where $p$, the degree of polarization, is the fraction of the radiation flux
that is polarized. When polarized emission passes through magnetized and ionised
regions, the local polarization angle $\psi$ (at position $\bmath{r}$) changes by
an amount depending on the wavelength $\lambda$ due to the Faraday effect
\[
\psi(\bmath{r})=\pso(\bmath{r})+\phi(\bmath{r}); \hspace{5mm}
	\phi(\bmath{r})=\lam^2 K\int^{\infty}_{z}\nel B_z \dif z^\prime,
\] 
where $K=0.81\rad\m^{-2}\cm^3\mkG^{-1}\p^{-1}$ is a constant,
$\nel$ is the number density of free thermal electrons,
$B_z$ is the component of the magnetic field along the line of sight (here
aligned with the $z$-axis), and the observer is located at $z\to\infty$.
$\phi(z)$ is known as the \emph{Faraday depth to a position $z$} and gives the change
in polarization angle of a photon of wavelength $\lam$ as it propagates from $z$
to the observer. The maximum amount of Faraday rotation in a given direction
is called the \emph{Faraday depth} \footnote{This terminology may
cause confusion: many authors, including \citet{Burn66} and \citet{Sokoloff98},
define the Faraday depth as $F/\lambda^2$, in our notation. However, it is more
convenient, and physically better motivated, to define the Faraday depth,
similarly to the optical depth, as a dimensionless quantity, as used by 
\citet{Spangler82} and \citet{Eilek89}}
\[
F=\phi(-\infty)=\lambda^2 K \int^{\infty}_{-\infty}\nel B_z \,\dif z^\prime\;.
\]
The observed amount of Faraday rotation, determined by the rotation measure 
$\RM=\dif\ps/\dif\lam^2$, cannot exceed $F$, i.e., $|\RM|\leq |F|\lambda^{-2}$. 

The value of $\RM$ is related to $F$, but often in a complicated manner
\citep[see, e.g.,][]{Burn66, Sokoloff98}. Simplest is the case of a Faraday
screen, where the source of synchrotron emission is located behind a
magneto-ionic region (e.g., because relativistic and thermal electrons occupy
disjoint regions): then $\RM=F\lambda^{-2}$. In a homogeneous region, where
relativistic and thermal electrons are uniformly mixed, $\RM=0.5
F\lambda^{-2}$.

Observations of linearly polarized emission provide the Stokes parameters $I$,
$Q$, $U$ which are related to $p$ and $\ps$ via $\cP=(Q+\mi U)/I$:
\begin{eqnarray}
p & = & \frac{(Q^2+U^2)^{1/2}}{I} \label{eq:qup}\;,\\
\ps & = & \label{eq:qupsi}
\sfrac{1}{2}\left[\arctan{\frac{U}{Q}}-\sfrac{1}{2}\pi(\sign Q-1)\sign U\right]\;,
\end{eqnarray}
and the polarized intensity is $P=(Q^2 + U^2)^{1/2}=pI$.
The complex polarization can be written in terms of $\phi$,
\begin{equation}
\cP = \frac{p_0}{I}\int_V W(\bmath{r}_\perp)\epsilon(\bmath{r})
        \exp\left\{2\mi\left[\psi_0(\bmath{r})+\phi(\bmath{r})\right]\right\}\dif V\;,
\label{cPF}
\end{equation}
where integration extends over the volume of the telescope beam $V$,
$W(\bmath{r}_\perp)$ defines the shape of the beam, a function of position in
the sky plane $\bmath{r}_\perp=(x,y)$, and $\epsilon(\bmath{r})$ is the
synchrotron emissivity. The total intensity is similarly given by $I=\int_V
W(\bmath{r}_\perp)\epsilon(\bmath{r})\,\dif V$; the Faraday depth is a function
of $\bmath{r}_\perp$, $F=F(\bmath{r}_\perp)$.

\section{Canals produced by a Faraday screen}
\label{sec:screen}

\begin{figure}
\begin{center}
\includegraphics[width=0.45\textwidth]{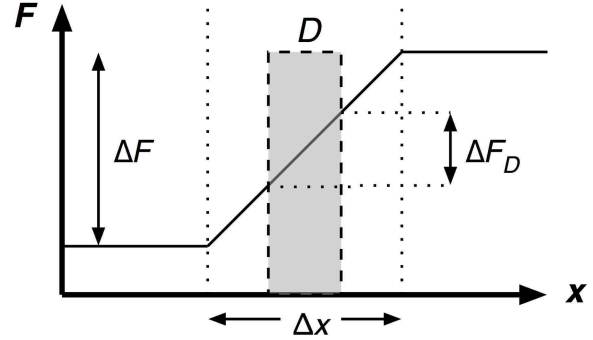}
\caption{Sketch showing the variation in Faraday depth $F$ with respect to
transverse distance in the sky plane $x$. The Faraday depth changes by $\DF$
across the distance $\Delta x$ and by $\DFD$ within the beamwidth $D$.}
\label{fig:grad}
\end{center}
\end{figure}

\begin{figure*}
\begin{center}
\includegraphics[width=.22\textwidth]{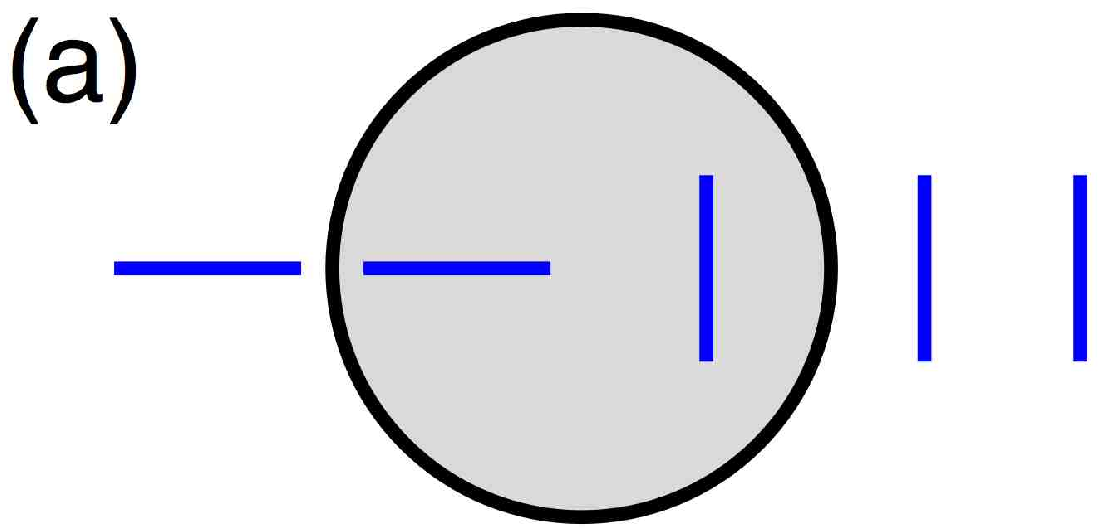}
\qquad
\includegraphics[width=.22\textwidth]{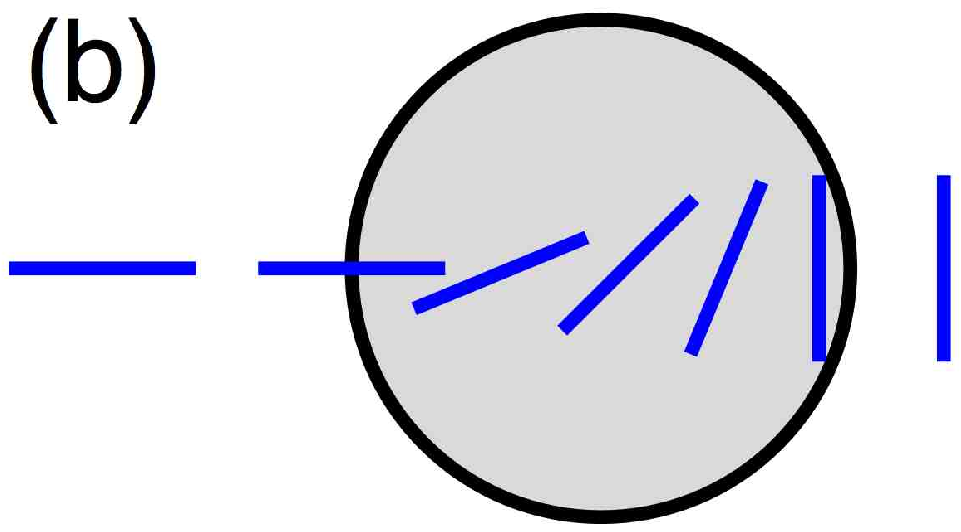}
\qquad
\includegraphics[width=.22\textwidth]{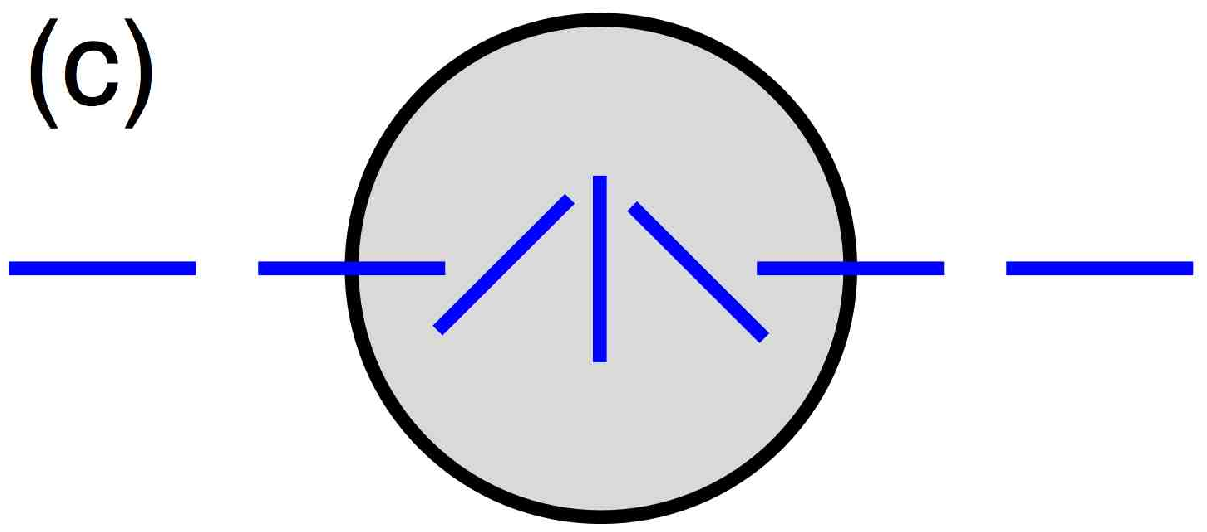}
\end{center}
\caption[]{An illustration of depolarization in a Faraday screen: the shaded
circle indicates the telescope beam, and dashes show the orientation of the
polarization plane at various positions. \textbf{(a)} Abrupt change of the
polarization angle $\ps$ by $90\degr$ in the middle of the telescope beam leads
to complete cancellation of polarization. This can be caused by a
\emph{discontinuous} change in $F$ by $90\degr$ in a Faraday screen.
\textbf{(b)} A similar but \emph{continuous} change of $\ps$ does not result in
strong depolarization. \textbf{(c)} A \emph{continuous} change in $\ps$ by
$180\degr$ can produce a canal. }
\label{screen}
\end{figure*}

Consider polarized emission from a uniform background source passing through a
Faraday screen -- a layer where no further emission occurs, but which rotates
the polarized plane. If $F$ varies with $\bmath{r}_\perp$, adjacent lines of
sight within the beam are subject to different amounts of Faraday rotation and
the observed degree of polarization decreases. If the variation of $F$ within
the beam, $\DFD$ (Fig.~\ref{fig:grad}), produces a $90\degr$ difference in
$\ps$, one might expect that the depolarization will be complete as illustrated
in Fig.~\ref{screen}(a), and a canal will be observed along contours defined by
$\DFD=(n+1/2)\pi$ with $n=0,1,2,\dots$.

The situation is more subtle, though. Figure~\ref{screen}(b) shows $\ps$
changing \emph{smoothly\/} by $90\degr$ across a beam, as a result of a
monotonic change in Faraday depth by the same amount, $\DFD=\pi/2$, as
in panel (a). It is obvious that polarized
emission does \emph{not\/} cancel within the beam; the polarized intensity will
be lower than for a uniform arrangement of angles but there will still be a
polarized signal detected with a polarization angle of about $-45\degr$.

\begin{figure}
\begin{center}
\includegraphics[width=0.45\textwidth]{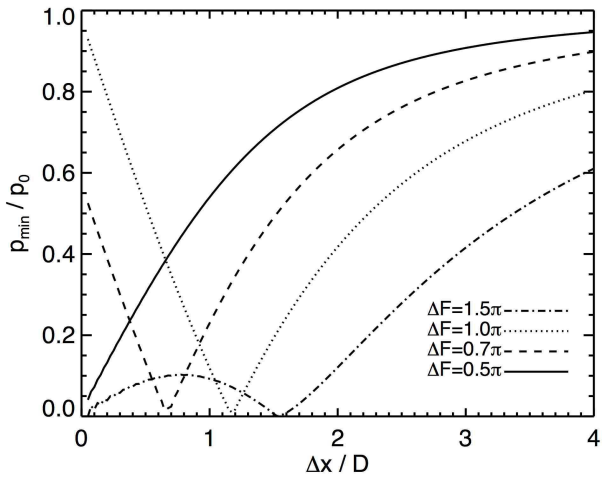}
\caption{The minimum degree of polarization in a canal produced by a gradient of
Faraday depth in a foreground screen, $p_\mathrm{min}$, as a function of the
relative linear extent of the gradient $\Delta x/D$, where $D$ is the FWHM of a
Gaussian beam and $x$ is the position in the sky-plane (see
Fig.~\ref{fig:grad}). Different curves represent different increments in Faraday
depth, $\DF$ (identified in the legend). For $\Delta x/D<1$ (an unresolved
gradient), $F$ changes over a distance smaller than the beamwidth and $\DF$ is
effectively discontinuous for $\Delta x/D\ll1$.}
\label{fig:pmin}
\end{center}
\end{figure}

Thus, gradients in $F$ across the beam can produce complete depolarization in a
Faraday screen if $F$ changes by $\DFD\simeq (n+1/2)\,\pi$ within a small fraction
of the beam width. As shown in Fig.~\ref{fig:pmin} (solid line), obtained using
Eqs~(\ref{eq:qup}) and (\ref{cPF}), the depth of a canal will only be less than
$10\%$ of the surrounding polarized emission if the gradient in $F$ occurs over
one fifth or less of the beam: $\Delta x/D\lesssim 0.2$, where $\Delta x$ is the
extent of the region where the gradient occurs and $D$ is the beamwidth.
\citet{Haverkorn04} reached a similar conclusion in an analysis of the depth and
profile of canals observed at $\lambda 84 \cm$ (see their  Sect.~4.4).

As illustrated in Figs.~\ref{screen}(c) and \ref{fig:pmin} (dotted line), a
variation in $F$ of a magnitude $\DFD=\pi$ across the FWHM of the Gaussian beam
(i.e. $\DF=\pi$ and $\Delta x/D\simeq 1$) can also produce strong depolarization
but now the change in polarization angle across the beam is $\Delta\ps\simeq
0\degr$. Similarly other gradients in $F$ can significantly reduce the degree of
polarization; for example $\DF=0.7\pi$ (dashed line in Fig.~\ref{fig:pmin})
across a region that is about $0.7\times$ the beamwidth will produce complete
depolarization and a change in angle of $54\degr$ across the beam.

Larger gradients in $F$ also result in strong depolarization and when
$\DF>\pi$ the degree of depolarization is less sensitive to the resolution. This
is illustrated in Fig.~\ref{fig:pmin} for the case where $\DF=1.5\pi$ (dash-dotted
line); complete depolarization occurs when $\Delta x/D=0,\,1.5$ but in the range
$0<\Delta x/D\lesssim 2$ the degree of polarization remains below the $10\%$
level. Thus an increment of $\DF=1.5\pi$, and more generally $\DF=(n+1/2)\pi$,
will produce one beam wide canals when the gradient occurs in a region roughly
equal to or less than the beamwidth, i.e. where $\Delta x/D\lesssim 1$. Other
large gradients in $F$, such as $\DF=n\pi$, depolarize to around
$10\%$ of the background $p$ and therefore can also
produce canals; however in this case one beam wide canals will only be formed 
when $\Delta x/D\simeq 1$.

So, if $F$ is a continuous, random, function of position (with sufficiently
strong fluctuations on the scale of the beam) we predict the existence of canals
with every change in polarization angle $0\degr<\Delta\ps<90\degr$ across the
beam. On the other hand \emph{a discontinuous change in $F$} by $(n+1/2)\pi$, 
will produce
canals across which the polarization angle changes by
$\Delta\ps=90\degr$. This insight is important as to-date all observed canals
have $\Delta\ps\simeq 90\degr$. The only other way in which Faraday
screens can consistently produce canals with this change in angle is if
\emph{all} the gradients in Faraday depth on the approximate scale of the
beamwidth are of the magnitude $\DF\simeq(n+1/2)\pi$, $n=1,2,\dots$ ($n=0$ in
this case only produces weak depolarization, as noted above); this is extremely
unlikely in a random medium.

If a system of observed canals has $\Delta\ps=90\degr$ and is caused by a
foreground Faraday screen --- see Section~\ref{sec:origin} for a diagnostic
test for the origin of canals --- then the distribution of rotation measure in
the screen must be discontinuous on the scale of the beam. \citet{Heitsch04}
used MHD simulations of Mach $10$ turbulence to show that $\RM$ gradients can be
steep enough to produce depolarized canals on smoothing with a sufficiently
large simulated beam. Note that a finite beam size is essential for the
production of canals by this mechanism; with perfect resolution the width of the
shock front will be resolved and no canal will appear. Furthermore, these canals
will not fill up if the data are smoothed (solid and dash-dotted curves for
$\Delta x/D\rightarrow 0$ in Fig.~\ref{fig:pmin}) whereas canals produced by
gradients other than $\DF=(n+1/2)\pi$ will disappear under smoothing at a rate
shown with dashed and dotted curves in Fig.~\ref{fig:pmin}.

\subsection{The mean separation of canals produced by shocks}
\label{sec:meansep}
An obvious source of Faraday depth discontinuities are shocks in the ISM. Then
the mean distance between the canals produced by a Faraday screen will be
related to the distance between the shock fronts. The expected separation of
suitable shock fronts can be derived using the model of interstellar
shock-wave turbulence of \citet{Bykov87} and then compared with the separation
of canals observed by \citet*{Haverkorn03}, claimed to arise in a Faraday
screen.

If the frequency distribution of interstellar shocks of Mach number $\M$ is
$\cG(\M)$, the mean number of shocks of a strength exceeding $\M$ that cross a
given position in the ISM per unit time can be obtained as
\begin{equation}
\label{eq:Fshock}
N(\M)=\int_{\M}^{\infty}\cG(\M)\,\dif\M\;,
\end{equation}
and the mean separation between the shocks in three dimensions follows as
\begin{equation}
\label{eq:Lshock3D}
L(\M)\simeq\frac{c_s}{N(\M)}\;,
\end{equation}
where $c_s$ is the sound speed. The separation in the plane of the sky is
reduced by  $\pi^{-1}\int^{\pi}_{0}\sin^2\theta\dif\theta=1/2$ due to projection
effects and multiplied by a further factor $L/d$ if the shocks occur in a region
of depth $d$. At an average distance to the shocks of $d/2$, the average angular
separation in the sky plane is then
\begin{equation}
\label{eq:Lshock2D}
\Lc\simeq\frac{L^2}{d^2}.
\end{equation}

For supernova-driven shocks, \citet{Bykov87} derived
\begin{equation}
\label{eq:GSNshock}
\cG(\M)=G_0[\M^{-(\alpha+1)}+3C(\alpha)f_\mathrm{cl}(\M-1)^{-4}]
\end{equation}
allowing for both primary and secondary shocks, where $\alpha$ is a numerical
factor ($\alpha=2$ for a supernova remnant in the Sedov phase, and $\alpha=4.5$
for a three-phase ISM); $C(\alpha)\approx2.3\times 10^{-2}$, $4.1\times 10^{-3}$
for $\alpha=2$, $4.5$ respectively; $f_\mathrm{cl}$ is the volume filling factor
of diffuse clouds that reflect the primary shocks to produce the secondary ones;
$G_0=\alpha S\frac{4}{3}\pi r_0^3$ with $S$ the supernova rate per unit volume
and $r_0$ the maximum radius of a primary shock. We then obtain the following
expression for the separation of shock fronts with $\M\ge\M_*$ in three
dimensions:
\begin{eqnarray}
\label{eq:SNshock}
L(\M) &\simeq & 9\p
        \left(\frac{c_s}{10\kms}\right)
        \left(\frac{\nu_\mathrm{SN}}{0.02\yr^{-1}}\right)^{-1} \nonumber \\
 & &\mbox{}\times \left(\frac{R}{15\kpc}\right)^{2}
        \left(\frac{h}{50\p}\right)
        \left(\frac{r_0}{100\p}\right)^{-3} \nonumber \\
 & &\mbox{}\times \left[\frac{1}{\M_*^{\alpha}}
        + \frac{\alpha C(\alpha) f_\mathrm{cl}}{(\M_*-1)^3} \right]^{-1},
\end{eqnarray}
where $\nu_\mathrm{SN}$ is the supernovae rate and $R$ and $h$ are the radius
and scale height of the star-forming disc. The term in square brackets is
approximately $1$ for $\alpha=4.5$ and $\M_*=1.2$.

If the magnetic field is frozen into the gas and the gas density increases by a
factor $\er$ at the shock then so will the field strength (we neglect a factor
$\sim 1/\sqrt{2}$ due to the relative alignment of the shock and field). A canal
forms where $F$ increases discontinuously by $\DFD=(n+1/2)\pi$. 
So the required density compression ratio is
\begin{equation}
\label{eq:epsilon}
\er\simeq \sqrt{\left|\frac{\DFD}{\meanb{F}}\right|+1},
\end{equation}
where $\meanb{F}$ is the mean value of $F$.
The gas compression ratio depends on the shock Mach number $\M$ as
\citep[e.g.,][]{Landau60}
\begin{equation}
\label{eq:mach}
\er=\frac{(\gamma + 1) \M^2}{(\gamma - 1) \M^2 + 2}\simeq4\frac{\M^2}{\M^2+3}\;,
\end{equation}
where we have used $\gamma\simeq 5/3$ for the ratio of specific heats, which
yields the value of the shock Mach number required to produce a canal:
\begin{eqnarray}
\label{eq:MRM}
\M_* & \simeq & \left(\frac{3\er}{4-\er}\right)^{1/2} \nonumber \\
     & = & \left(\frac{4}{3(|\DFD/\meanb{F}|+1)^{1/2}} 
        - \frac{1}{3}\right)^{-1/2}.
\end{eqnarray}

In the field of canals observed by \citet{Haverkorn03} at $\lambda 84\cm$,
$\meanb{\RM}\simeq -3.4\radmm$ and a by-eye estimate of their mean separation
gives $\Lc\simeq 45'$. The most abundant canals are produced by the shocks which
can generate $\DFD=\pi/2$ and from Eq.~(\ref{eq:MRM}) this requires $\M_*\simeq
1.2$. Using $f_\mathrm{cl}=0.25$, taking $\alpha=4.5$ and the parameter values
used to normalize Eq.~(\ref{eq:SNshock}), we obtain $L\simeq 10\p$ for shocks
with $\M\ge\M_*$. Our estimate of $L$ includes shocks with $\M>\M_*$ that will
not produce canals, but the strong dependence of $\cG$ in
Eq.~(\ref{eq:GSNshock}) on Mach number means that $L$ will underestimate the
separation of canal generating shocks insignificantly. Using
Eq.~(\ref{eq:Lshock2D}), with $L\simeq 10\p$ and $\Lc\simeq 45'$, we find that
the canals observed by \citet{Haverkorn03} are compatible with a system of
shocks occurring in a Faraday screen with a depth of $d\simeq 100\p$. The
maximum distance in this field, beyond which emission is completely depolarized,
is estimated to be $\sim 600\p$ by \citet{Haverkorn03}. Thus this model for the
canals' origin requires that the nearest $100\p$ in the direction
$(l,b)=(161\degr,\,16\degr)$ is effectively devoid of cosmic ray electrons; if the
cosmic rays normally spread over a distance $v_\mathrm{A}t\sim 1\kpc$, where
$v_\mathrm{A}\simeq 20\kms$ is the Alfv\'en speed and $t\simeq 3\times 10^7\yr$
their lifetime, such a condition is difficult to explain.

\section{Canals produced by differential Faraday rotation}
\label{sec:diffrot}

In Section~\ref{sec:screen} we discussed the depolarization effects of
Faraday rotation measure gradients transverse to the line of sight acting on
smooth polarized background emission. Now we will consider a uniform layer in
which both emission and Faraday rotation occur. This gives rise to the well
known effect of differential Faraday rotation \citep{Burn66, Sokoloff98},
where polarized emission from two positions along the line of sight separated by a
rotation of $\Delta\phi=\pi/2$ exactly cancel, thus reducing the degree of
polarization. When a line of sight has a total rotation of $F=n\pi$ there is
total cancellation of all polarized emission in the layer and the degree of
polarization is $p=0$. Canals produced by this mechanism are discussed by
\citet{Shukurov03}.

\subsection{Differential Faraday rotation in a non-uniform medium}
\label{subsec:diffrot}

We now investigate how canals produced by differential Faraday rotation are
affected by deviations from uniformity in the magneto-ionic medium. First we
will discuss the case of a two-layer medium in which the synchrotron emissivity
is different in each layer. Then we allow for random fluctuations of $F$
(the latter case produces what is known as Faraday dispersion).

For a two-layer medium the fractional polarization can be written as
\begin{eqnarray}
\label{eq:twolayer}
p^2 & = & \frac{\po^2}{I^2}\left(I\one^2\frac{\sin^2 F\one}{F\one^2}
		+ I\two^2\frac{\sin^2 F\two}{F\two^2} \right. \nonumber \\
  & & \left. + 2 I\one I\two \frac{\sin F\one \sin F\two}{F\one F\two} \cos F \right) 
\end{eqnarray}
where we have assumed $\pso=0$, $I\one$ and $I\two$ are the synchrotron fluxes
from the furthest and nearest layers to the observer respectively,
$I=I\one+I\two$, and $F\one$ and $F\two$ are the Faraday depths through the two
layers. [This is similar to Eq.~(10) in \citet{Sokoloff98}, but with typos
corrected.] We assume that $F=F\one+F\two=\pi$, the condition for canals to form
due to differential Faraday rotation, and investigate what will happen to the
canals if $I\one\ne I\two$. We parametrize the difference in the synchrotron
emissivity in the two layers as $\epsilon=(I\two-I\one)/I\one$, choose equal
Faraday rotation through each layer for simplicity so $F\one=F\two=\pi/2$, and
substitute these into Eq.~(\ref{eq:twolayer}). We then obtain the dependence of
the minimum degree of polarization on $\epsilon$:
\begin{equation}
\label{eq:twolayerfill}
\frac{p}{\po}=\frac{|\epsilon|}{\pi}\frac{1}{(1+\epsilon/2)^2} 
	\simeq \frac{|\epsilon|}{\pi}
	- \mathcal{O}(\epsilon^2).
\end{equation}
Thus, for a
moderate difference in emissivity of $\epsilon = 1/2$ between two layers
the canals will have a depth of $p/\po\simeq 1/6$ compared to $p/\po = 0$ for
a strictly uniform medium; these canals will still be clearly visible and can
be interpreted as e.g. contours of $F$ or $\RM$ \citep{Shukurov03}. Where the
difference in emissivity is greater, say $\epsilon\sim 2$ or more as one might
expect viewing distant  bright spiral arm emission near the Galactic plane,
the canals will more readily fill up, become much less distinct and more
difficult to interpret confidently.

Now let us consider the effect of random fluctuations in Faraday rotation,
sometimes called internal Faraday dispersion. We start with Eq.~(34) of
\citet{Sokoloff98}:
\begin{equation}
\label{eq:disp}
\cP=\po\frac{1-\exp(-S)}{S},
\end{equation}
where $S=2\sigma_{F}^2 - 2iF$ and $\sigma_{F}$ is the standard deviation of the
Faraday depth $F$. In order to produce canals we need $F=\pi$ and then
\begin{equation}
\label{eq:dispfill}
\frac{p}{\po} \simeq
	\frac{\sigma_{F}^2}{\pi}\left( 1-\frac{1}{2}\frac{\sigma_{F}^4}{\pi^2}\right),
	\hspace{5mm}\sigma_F\ll 1.
\end{equation}
The term in brackets shows the powerful depolarization resulting from internal
Faraday dispersion (see \citet{Sokoloff98}): for large enough $\sigma_{F}$
depolarization is complete. However, as long as $\sigma_{F}<1$, canals produced
by differential Faraday rotation will only fill up by $1/3$ or less. For
example, at $\lam 20\cm$ the canals will not be destroyed as long as the
dispersion in Faraday rotation measure is less than $30\radmm$. This is why
canals are still visible in the $\lam 20\cm$ map studied by \citet{Shukurov03}
which has a dispersion in rotation measure of about $\sigma_{\RM}\simeq
10\radmm$, i.e. $\sigma_{F}=\sigma_{\RM}\lam^2\simeq 0.4$.

\section{Observational diagnostics for the origin of a canal}
\label{sec:origin}

The complex polarization emitted by a layer producing differential Faraday
rotation can be written as \citep{Burn66}
\begin{equation}
\label{eq:diffrot}
\cP = \po\frac{\sin F}{F} \me^{2\mi(\pso + F/2)}.
\end{equation}
We can see immediately that if $\pso=0$ we have $Q/I=\mathrm{Re}\,\cP\propto \sin
F\cos F$ and $U/I=\mathrm{Im}\,\cP\propto \sin^2 F$. Since the orientation of the
coordinate system in which $Q$ and $U$ are defined is arbitrary (i.e. the
reference line from which we measure polarization angles can have any orientation), we can
always choose a system in which $\pso=0$ near a canal. At the axis of a
canal $Q=U=0$, but in the case of a canal produced by differential Faraday
rotation there exists a reference frame in the ($Q$, $U$) plane where one of the
two Stokes parameters changes sign across a canal but the other does not.

If, otherwise, a canal is produced by discontinuities in Faraday
rotation across the beam (Sect.~\ref{sec:screen}) we have \citep{Fletcher06a}
\begin{equation}
\label{eq:quscreen}
\cP=\frac{Q+\mi U}{I} =  \po\frac{\sin \DFD}{\DFD} \me^{2\mi(\pso+F)},
\end{equation}
in the vicinity of a canal and both $Q$ and $U$ will change sign across a canal
(given $\pso=0$) except in the special case $\DFD=F+2n\pi$. The change in sign
occurs since $Q/I=\po\cos[2(\pso+F)]$ on one side of the canal and
$Q/I=\po\cos[2(\pso+F)+\pi]$ on the other: a similar variation occurs in $U/I$.
Therefore the behaviour of the observed Stokes parameters $Q$ and $U$ across a
canal provides a way to distinguish between canals produced by foreground
Faraday screens and those resulting from differential Faraday rotation.

%
\section{Summary}
\label{sect:summary}

The main points of this paper can be summarized as follows:
\begin{enumerate}
\item The behaviour of the Stokes parameters $Q$ and $U$ in the vicinity of a
canal allows one to identify whether a foreground Faraday screen or differential
Faraday rotation is the cause of the canal (Section~\ref{sec:origin}).

\item A foreground Faraday screen can produce canals with any polarization angle
change across the canal, $0<\Delta\ps<90\degr$. However, discontinuous jumps
in the Faraday depth will only produce canals with $\Delta\ps\simeq 90\degr$
(Section~\ref{sec:screen}). 

\item If shocks produce the discontinuities in a foreground Faraday
screen that generate canals, the mean separation of the canals can provide
information about the Mach number and separation of shocks in the screen
(Section~\ref{sec:meansep}).

\item Canals produced by differential Faraday rotation are sensitive to
non-uniformity in the medium along the line of sight, systematic or random.
However, they will remain recognisable if the synchrotron emissivity varies by
less than a factor of about  $2$ or if the standard deviation of the Faraday depth
is $\sigma_{F}<1$ (Section~\ref{subsec:diffrot}).
\end{enumerate}

\section*{Acknowledgments}
This work was supported by the Leverhulme Trust under research grant
F/00~125/N.


\bsp

\label{lastpage}

\end{document}